\documentclass[conference]{IEEEtran}
\IEEEoverridecommandlockouts
\usepackage{subfig}
\usepackage{multirow}
\usepackage{cite}
\usepackage{amsmath,amssymb,amsfonts}
\usepackage{algorithm}
\usepackage[noend]{algpseudocode}
\usepackage{graphicx}
\usepackage{textcomp}
\usepackage{xcolor}
\def\BibTeX{{\rm B\kern-.05em{\sc i\kern-.025em b}\kern-.08em
    T\kern-.1667em\lower.7ex\hbox{E}\kern-.125emX}}

\graphicspath{{./Images/}}

\begin{document}

\title{ragamAI: A Network Based Recommender System to Arrange a Indian Classical Music Concert
}

\author{\IEEEauthorblockN{Arunkumar Bagavathi\IEEEauthorrefmark{1}, Siddharth Krishnan\IEEEauthorrefmark{2}, Sanjay Subrahmanyan, S.L. Narasimhan}
\IEEEauthorblockA{\textit{Department of Computer Science}\IEEEauthorrefmark{1}\IEEEauthorrefmark{2} \\
\textit{Oklahoma State University\IEEEauthorrefmark{1}, University of North Carolina at Charlotte\IEEEauthorrefmark{2}}\\
abagava@okstate.edu, skrishnan@uncc.edu, sanjaysub@gmail.com, yessel@gmail.com}
}

\maketitle

\begin{abstract}
South Indian classical music (\emph{Carnatic music}) is best consumed through live concerts. A carnatic recital requires meticulous planning accounting for several parameters like the performers' repertoire, composition variety, musical versatility, thematic structure, the recital's arrangement, etc. to ensure that the audience have a comprehensive listening experience. In this work, we present \emph{ragamAI} a novel machine learning framework that utilizes the tonic nuances and musical structures in the carnatic music to generate a concert recital that  melodically captures the entire range in an octave. Utilizing the underlying idea of playlist and session-based recommender models, the proposed model studies the mathematical structure present in past concerts and recommends relevant items for the playlist/concert. \emph{ragamAI} ensembles recommendations given by multiple models to learn user idea and past preference of sequences in concerts to extract recommendations. Our experiments on a vast collection of concert show that our model performs 25\%-50\% better than baseline models. \emph{ragamAI}'s applications are two-fold. 1) it will assist musicians to customize their performance with the necessary variety required to sustain the interest of the audience for the entirety of the concert 2) it will generate carefully curated lists of south Indian classical music so that the listener can discover the wide range of melody that the musical system can offer.

\end{abstract}

\begin{IEEEkeywords}
recommender system, music information retrieval, tone embeddings
\end{IEEEkeywords}
\section{Introduction}

    


The analysis of classical music has become a mainstay in the field of music information retrieval (MIR) over the last decade. Recent research has investigated several nuances of classical music, be it in rhythm cycles, tempo estimation, beat tracking, instrument classification, and melodic analysis, using data-driven techniques. However, there is a paucity of work in bridging these information retrieval techniques to applications like playlist recommendations and automated concert planners. In this work, using the lens of south Indian classical music, also known as \emph{Carnatic} music, we exploit the grammatical structure and mathematical underpinnings of the music system to develop a machine learning model that can be applied to build a concert plan. While there are several recent works that model different aspects of Indian classical music, to the best of our knowledge we are the first to use the melodic structure, defined as a \emph{ragam}, present in the music system to design a recommender system. Carnatic musicians have been relying on their repertoire and memory to design song lists in the past. An automated system will enhance the experience and make it more versatile in the development of a concert song list. The proposed model can also be used as an ideal tool to improve a repertoire given that it can draw on source material from different types of databases that suit the case.  Moreover, the model can also be used to generate music playlists for streaming applications like Spotify, Pandora, last.fm, etc. to provide an enriching experience that encompasses the musical range offered by the carnatic system in a song sequence.

The efficacy of a session-based recommendation model is to provide desired items for a user in their current session based on their past preferences. For example, a video streaming service like YouTube predicting the user's preference to watch next. In the research literature, there are several frameworks and methodologies for music and video playlist recommendations. Given the mathematical details in south Indian classical music and the task of arranging songs in a concert for a musician, we formulate this problem as session (concert)-based playlist (song) recommendation and propose a network based recommender model. We give a network/graph representation for sequences in Indian classical music concerts with our proposed \emph{Raaga network}. Along with features extracted from the network, the proposed model capture co-occurrences, melodic patterns, and musician's preference of flow in concerts as features. With experiments we show that inclusion of such features improve the performance of recommender model in constructing Indian classical music concerts.

In this work, we aim to solve following research questions by providing a music recommender framework for Indian classical music concerts:

\begin{itemize}
	\item \textbf{How the concerts can be structured to understand its organization?} We propose a network structure called \emph{Raaga Network} that captures order, co-occurrence, and context of each item(song/raaga) in concerts.
	\item \textbf{Can historical data help musicians to build new repertoire?} We postulate a network-based machine learning model to recommend items(songs/raagas) based on their underlying structure and mathematical constructs in Raaga Network. We frame this model to extract a set of recommendations for a given sequence of songs.
	\item \textbf{How useful are item(songs/raagas) recommendations given by the models?} We evaluate the usefulness of recommendations from the proposed framework using offline methods like \textit{precision} and \textit{discounted cumulative gain}.
\end{itemize}
\section{Related Work}

Innovations in \emph{machine learning} helped the scientific community to contribute to large scale recommendation systems like Pinterest~\cite{you2019hierarchical} and Spotify~\cite{Pichl2017Spotify}. All recommender systems fall in one of the three categories: content-based (\emph{domain dependent}), collaborative (\emph{domain independent}), and hybrid (\emph{multi-model frameworks})~\cite{isinkaye2015recommendation}. These algorithmically constructed systems are evaluated using either offline measures like precision, recall, mean average precision(MAP), Normalized Discounted Cumulative Gain(nDCG) or online measures like A/B test and p-value~\cite{gunawardana2015evaluating}.

Advancements in recommendation systems have been applied to music domain as well, Spotify and Pandora for example. The most popular problem in music recommendations is to select a set of songs as a playlist for a user based on their mood and preferred genre and artist~\cite{Bonnin2014automated}. This problem has been answered using multiple methods like: frequent pattern mining~\cite{chen2012playlist}, collaborative filtering~\cite{Cheng2017exploring}, and hybrid models~\cite{hornung2013evaluating}. IN addition to the playlist recommendation models are the session-based prediction models (i.e.), ~\cite{Hidasi2016SessionbasedRW,Jannach2017rnn,hidasi2018recurrent}. \emph{Novelty} and \emph{diversity} are considered to be important evaluation measures for music recommendation models~\cite{schedl2015tailoring} to provide the user interesting and unexplored suggestions. In this paper, we present a session-based recommendation model that utilizes deep learning algorithms to give recommendations.

Machine learning has been used in the Indian classical music over the past decade to study multitude of concepts. For example, classifying recordings from YouTube based on the swara using random forest algorithm~\cite{dighe2013swara}, using pitch information in music signals~\cite{kumar2014identifying}, and a vector based classification model~\cite{gulati2016phrase}, similar to text classification model. Identifying tonic of a music from multi-pitch analysis of the given audio was also given as a classification problem~\cite{salamon2012multipitch}. Unlike the existing methods, we describe methods through the lens of recommender systems. We incorporate the idea of embedding possible features from historical concerts and frequent pattern mining methods to make recommendations. Moreover, our model is designed as a human computer interaction system to communicate recommendations.


\section{Background}

Carnatic music is the classical music system of Southern India. The Carnatic music tradition is built on the melodic foundation of the ragam or scale that encompasses a collection of swaras (notes) in an octave. The scale follows a 12 note per octave system. This has been expanded into a 16 note system through the 72 melakarta scheme. The ragam system over the last 3 or more centuries has evolved into being categorized according to the mela janya scheme where by ragams are either parent scales or derivatives (mela or janya).

A melakarta by definition will have the same notes in both its ascending and descending scales. The math in the organization of the melakarta scheme will give an idea of the variety of ragams available for musicians to learn, practice, and perform. A typical octave will have the 7 basic notes: S R G M P D N. The S and the P are fixed and static notes. R, G, M, D, N will have different variations lie R1, R2, R3, G1, G2, G3, M1, M2, D1, D2, D3, , N1, N2, N3. If one looks at the frequency values of these notes, R3 corresponds to G1 and D3 corresponds to N1. The melakarta scheme by definition will have a set of 7 notes with S and P constant. So the first ragam will be S R1, G1, M1, P, D1, N1, . The scheme also follows the rule that there will only be one variety of each note that is permissible. So there can be no occurrence of R1 and R2 in a melakarta. Another constraint is that since the frequency values of R3 and G1 are the same they will never occur together. So typically the scheme follows the system of the ragams $1-36$ having S, M1 \& P as constants and $37-72$ having S, M2 \& P as constants.

Subdividing the ragams further the first 6 will have S R1, G1, M1, P as constants and the six variations will add the combinations D1/N1, D1/N2, D1/N3, D2/N2, D2/N3 \& D3/N3. The next six will have R1/G2 with the six varities of D/N and so on to arrive at a total of 72 ragams.

The janya ragams are derivative scales from the parent. Janya ragams have no rules. One can miss few notes in the scale and they can even have different ascending and descending scales. 

A typical Carnatic music concert consists of pieces that are performed over a $60 - 240$ minute period depending on the artist and location. Each composition in a concert can be of varying lengths and usually revolving around a central piece (main ragam). Performers therefore have to train a huge repertoire of compositions set to perform in a variety of ragams. These ragams offer melodic variety in terms of sound, color, and aesthetics. This variety comes through a process of inherent differences because of the specific notes occurring as well as musical and aesthetic differences based on how they are handled by the musician. 


The challenge therefore to the performer is to come up with a list of pieces that can offer variety musically, keep the listener engaged and remain fresh so as to avoid being stale and monotonous. A primary constraint of any song list would be non repetition of ragams. The Carnatic music system technically has innumerable ragams, however the numbers in vogue and practiced on a regular basis would probably be in the region of less than 1000. The second aspect of choosing pieces will be compositions. The process involved is coming up with an ideal list of pieces for a concert is taking into consideration both compositions and ragas. All compositions are set to a particular ragam and the entire rendition of the same will conform to that scale of notes only. We plan to exploit composition based recommendations in our follow-up work.
\section{Methodologies}
\begin{figure}
\centering
\includegraphics[scale=0.35]{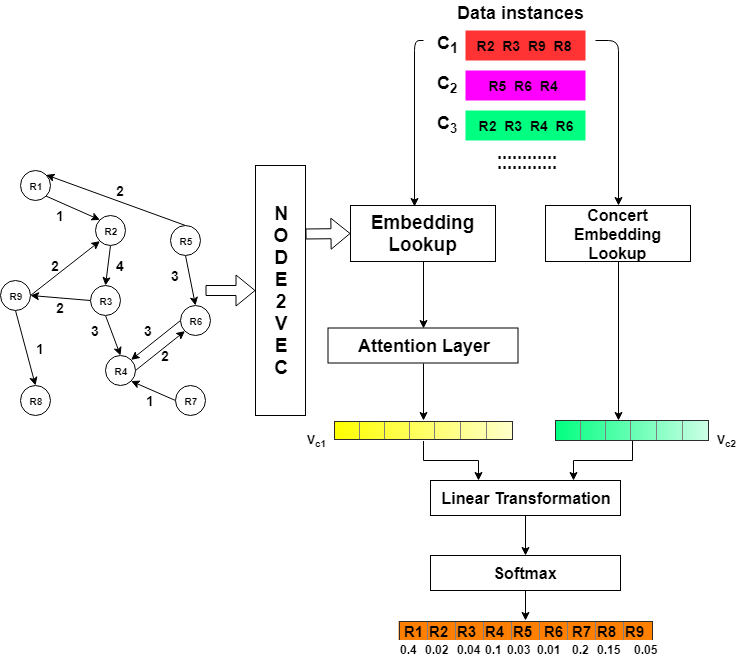}
\caption{Architecture of the ragamAI. The model composes 3 layers: 1. Network embedding(skip-gram model) and Concert embedding(embed hand picked features), 2. Vector aggregation from two embedding, and 3. Recommendation} \label{fig:ensembel_model}
\end{figure}

The proposed ragamAI framework relies on the influence of two separate models: 1. a deep attention model to capture the importance of sequence of ragams, and 2. an embedding model to capture the importance of hand picked features to train. Unlike other methods, which predict next event or item in a given sequence~\cite{chen2012playlist,Hidasi2016SessionbasedRW,hu2017diversifying,wang2018attention,wu2019session}, the proposed model(s) predict all forthcoming items (songs) in a given concert.  All our models use both 1-to-1 mapping of features and feature similarity to measure relevancy of recommendations. An overview of the proposed model is represented in Figure~\ref{fig:ensembel_model}


\subsection{Raaga network}

\begin{figure}
\centering
\includegraphics[scale=0.5]{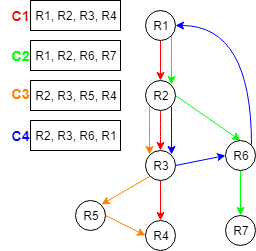}
\caption{Sample raaga network constructed from concerts. Concerts \textbf{C1},\textbf{C2},\textbf{C3}, and \textbf{C4} are represented as an ordered list of ragams. Colors in the figure are replaced with edge weights in all experiments. \textbf{NOTE:} Each song in a concert consists of only one ragam} \label{fig:raaga_network}
\end{figure}

Networks or graphs have been considered as a promising framework to study variety of applications like influence modeling~\cite{aslay2018influence}, community detection~\cite{chien2018community}, and recommender systems~\cite{ying2018graph}. Their organization of nodes and edges help to study the structural organization and positional values of entities (nodes and communities). We propose the \emph{Raaga network} as a systematic representation for concerts, where each concert is an ordered transaction of items (songs/ragams). Formally, we define the \emph{Raaga network} as $\mathcal{G}=(\mathcal{R},\mathcal{E},\textit{w})$, where $\mathcal{R}$ is a set of items/ragams, $\mathcal{E}$ is a set of edges representing the flow from one ragam($\mathcal{R}_m$) to the next($\mathcal{R}_n$) $\forall \mathcal{R}_m,\mathcal{R}_n \in \mathcal{R}$, $\textit{w}$ is edge weight representing the overall frequency of the sequence, and $|\mathcal{R}|=$ Total number of unique items/ragams. Example raaga networks are given in Figures~\ref{fig:ensembel_model} and \ref{fig:raaga_network}. We capture the order of ragams in concerts by maintaining a set of attributes $\mathcal{A}_i={a_1,a_2 \dots a_n}$, where $i=\mathcal{R}_1,\mathcal{R}_2 \dots \mathcal{R}_{|\mathcal{R}|}$. Example attributes include concert id, ragam type, and tonic information of a ragam. The proposed network gives intuition on the flow of each concert and gives an intuition of a ragam's context throughout all concerts. We use this network as an underlying framework to produce recommendations in the proposed model.

\subsection{Ragam representation learning using node2vec}
The structural knowledge from the ragam network can help the recommender model to give better results to the user based on their co-occurrence. We use a network embedding model: \textbf{node2vec}~\cite{grover2016node2vec} to extract features of nodes in the raaga network. This model capture the node context with \textit{k}-iterations of random walks of length \textit{l} along with the optimization function given in Equation~\ref{eq:node2vec_optimization}

\begin{equation}
\label{eq:node2vec_optimization}
\max_f \sum_{v \in \mathcal{R}} log P(N(v)|f(v))
\end{equation}

	where $N(v)$ is the neighbor nodes of node $v$ and the likelihood of the neighborhood of a node is modeled as a softmax function given in Equation~\ref{eq:node2vec_likelihood}
	
\begin{equation}
\label{eq:node2vec_likelihood}
P(N(v)|f(v)) = \prod_{m \in N(v)} \frac{exp(f(m).f(v))}{\sum_{w \in \mathcal{R} exp(f(w).f(v))}}
\end{equation}

Thus with node2vec, we collect feature representation of each ragam from all the concerts. Since each data instance is given as a set of ragams in a concert, we aggregate the ragam vectors. Like many other session based applications, the number of songs in every music concerts also vary. Our model handles such variable length inputs by padding zero vector(s). Since order in the song sequence given in the feature vector may have different priorities, we use an attention model to get the concert vector ($V_{c_1}$) (given in Equation~\ref{eq:attention})

\begin{equation}
\label{eq:attention}
V_{c_1} = \sum_{i=1}^{k} \sigma(W_1 s_{i-1} + W_2 s_i + c) s_i
\end{equation}

 where $W_1,W_2 \in \mathbb{R}^{d*d}$, $k$ is the number of songs in the concert, and $s$ is the vector representation of the given ragam $i$.

\subsection{Representing the hand picked features}


Instead of optimizing the learning model (like node2vec), we give a simple strategy to embed a concert (or a set of songs/raaga) into a vector space. In other words, we represent a concert a distribution of concert features (\textit{audava},\textit{shadava}, and \textit{sampoorna} for example). All features considered for this study is given in Table~\ref{tab:features}. For a given concert, we create one-hot encoded vectors for each ragam based on these hand picked features. We aggregate the features with element-wise average on all the vectors. Thus we represent entire concert into a vector space($V_{c_2}$). 


\subsection{RagamAI model}
With concert vectors from the attention model ($V_{c_1}$) and the concert embedding model ($V_{c_2}$), we perform a linear transformation after concatenating the vectors to obtain the aggregate vector using the Equation~\ref{eq:aggregate}

\begin{equation}
\label{eq:aggregate} 
V_g = W_3(V_{c_1} \oplus V_{c_2})
\end{equation}

We apply softmax function to get the $|\mathcal{R}|-dimensional$ output vector ($y'$), where each element in $y'$ vector represent a score for each ragam for recommendations and $|\mathcal{R}|$ is total number of nodes in the raaga network $\mathcal{G}$. 

We use cross-entropy, given in Equation~\ref{eq:loss}, as a loss function for training the proposed model.
\begin{equation}
\label{eq:loss}
\mathcal{L}(y') = -\sum_{i=1}^{|\mathcal{R}|} y_i log(y_i') + (1-y_i) log(1-y_i)
\end{equation}

\section{Experiments and Results}

\subsection{Dataset}
\begin{figure*}[!ht]
	\centering	
		\subfloat[Overall distribution of songs across concerts][]{
			\includegraphics[width=0.43\textwidth]{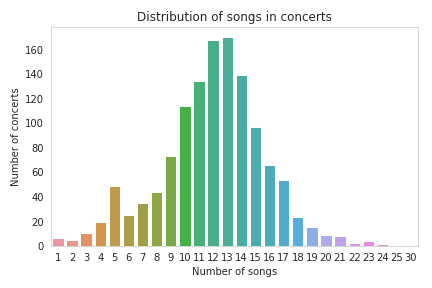}
			\label{song_dist}
		}
		\hfill
		\subfloat[Distribution of janya and melakartha ragams across concerts][]{
			\includegraphics[width=0.43\textwidth]{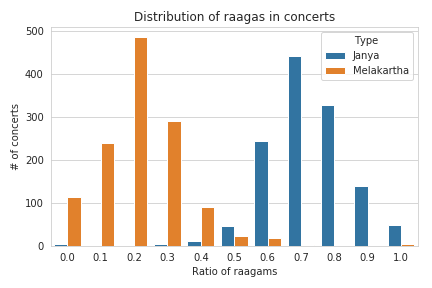}
			\label{janya_dist}
		}
	
		\qquad	
	
		\subfloat[Distribution of melakartha ragams across concerts][]{
			\includegraphics[width=0.43\textwidth]{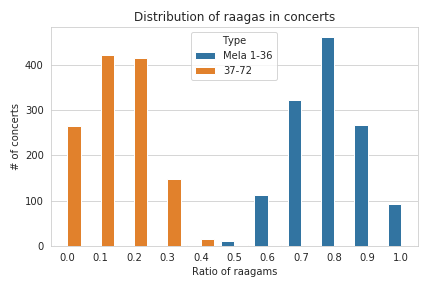}
			\label{mela_dist}
		}
		\hfill
		\subfloat[Distribution of ragam features across concerts][]{
			\includegraphics[width=0.43\textwidth]{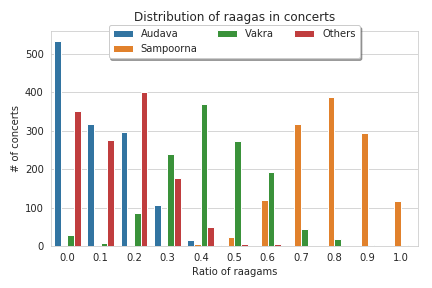}
			\label{feature_dist}
		}

		\caption{Distribution of various aspects(features) of south Indian classical music performed by one musician.}
		\label{fig:data_dist}
\end{figure*}

We use the real world data consisting of concerts performed by multiple Indian classical musicians. The data comprises of \textit{1664} concerts performed between 1984 and 2019 with more than \textit{800} unique songs and \textit{280} unique ragams. Each concert is a set of songs, where each song can map to its corresponding ragam. A performer's creativity, impromptu improvisations, and different song durations during the concert, the number of songs in each concert is a dynamic value. Also, since all  these concerts are performed in several locations around the world, the songs and the concert style (organization of songs and ragams) are unique. In Figure~\ref{fig:data_dist}, we give basic data distribution available in our data. As implied in Figure~\ref{song_dist}, most concerts have in the range of 9-16 songs. Figure~~\ref{janya_dist},\ref{mela_dist}, and \ref{feature_dist} give distribution of other features in concerts and we give proportion of songs matching these features in concerts. For example, Figure~\ref{janya_dist} gives the ratio of number of songs in a concert is \textit{janya} or \textit{melakartha} ragam, and Figure~\ref{feature_dist} gives the ratio of types of ragams in concerts(\textit{audava},\textit{sampoorna},\textit{vakra}, and other ragams). Due to space constraints, we neglected the distribution of rarely occurring type of ragam(\textit{shadava} ragam) from this figure. Our data also consists of metadata of all 280 ragams. We use concert structure(organization of ragams in each concert) to build the raaga network and we populate metadata as node features in the network. We further use this network of ragam patterns in experiments for the proposed model to recommend ragams.


\begin{table}[!t]
\caption{Features of ragams in the RagaNetwork. Each node in the raga network depicts a ragam and thus each node holds the below features. Although position of ragams in a concert is important, it can be identified by traversing before and after nodes matching the concert id.} \centering
\label{tab:features}
\centering

\begin{tabular}{|p{2cm}||p{5.8cm}|}
\hline
\textbf{Feature} & Description \\
\hline
Janya/Melakartha & Boolean feature on weather a ragam is janya \\
\hline
Mela number & Melakartha number of a ragam \\
\hline
Mela category & Melakartha category of a ragam \\
\hline
Type & Ragam type - Audava or Shadava or Sampoorna ragam \\
\hline
Combo & Some ragams fall between 2 different ragam types \\
\hline
Vakram & Boolean feature on whether a ragam is Vakra ragam \\
\hline
Concert & List of concert ids that the ragam is present \\
\hline
\end{tabular}
\end{table}



\subsection{Baseline Models}
Below are some of the popular baseline models that we used to compare our methods

\begin{itemize}
	
	\item \textbf{Item-kNN}~\cite{linden2003amazon}: Item-kNN is similar to extracting the nearby recommendations. In this method, we find co-occurring items normalized by popularity of all items in the list.
	
	\item \textbf{FPMC}~\cite{rendle2010factorizing}: \textit{Factorized Personalized Markov Chain} is one of the popular models to recommend sequential item in a list. \textit{FPMC} models the user preference on a list and the preference transitions to recommend next set of items. 
	
	\item \textbf{SWIWO}~\cite{hu2017diversifying}: Session-based Wide In Wide Out(SWIWO) is a deep learning based recommender model that uses a both user and item based feature set to predict the next item. To adapt to our problem, we eliminate the user feature section and add a softmax layer to predict the score of items for a given sequence. 
\end{itemize}

We compare proposed model against existing methods in producing recommendations for south Indian classical music concerts. To make fair comparison, we use offline evaluation methods for all algorithms, similar to comparisons made in existing works. In particular, we use following evaluation methods:

\begin{itemize}
	\item Precision@k: We evaluate on number of recommendations ($k$) to make (i.e.) We evaluate from $k=1$ (Recommending just one item) to $k=15$ (Recommending 15 items).
	\item Normalized Discounted Cumulative  Gain(nDCG): It is a measure to evaluate the rank of relevant recommendations, calculated using the formula given in Equation~\ref{eq:ndcg}
	
\begin{equation}
\label{eq:ndcg}
	nDCG = \frac{\sum_{i=1}^{n} \frac{1}{log_2(i)}}{DCG_{id}}
\end{equation}
	 
, where $n$ is the number of items in the list to predicted correctly, $i$ is the position of the relevant item in the recommendation, $DCG_{id}$ is an ideal score for the given test sample. Similar to Precision@k, we give results of $nDCG@k$, where $k$ is the number of items to recommend.
\end{itemize}

\subsection{Results}

\begin{figure}
\centering
\includegraphics[scale=0.5]{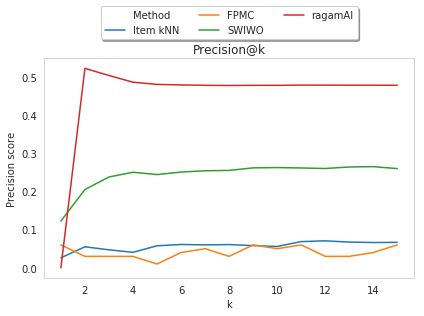}
\caption{Precision@k evaluation for different models} \label{fig:precision}
\end{figure}

\begin{figure}
\centering
\includegraphics[scale=0.5]{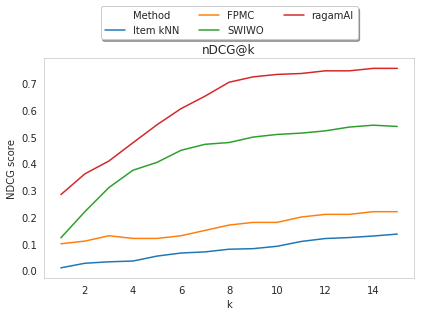}
\caption{nDCG@k evaluation for different models} \label{fig:NDCG}
\end{figure}

For all our experiments we split the available concerts into training (80\%) and test (20\%) data, where the training data is used as historical data and the test data is considered as new concerts. Since the goal is to recommend ragams for incomplete concert list and number of ragams in a concert is inconsistent, we conduct our experiments with an assumption that the user starts with one ragam (first ragam from the test concert), the system gives prediction and evaluate obtain results, then it considers first two ragams, and the cycle continues. For training the model, we split each concert into multiple sequences in such a way that all instances have at least one user input ragam and one prediction. All results that we report here is an average score of all measures.


Figure~\ref{fig:precision} gives precision@k scores and the Figure~\ref{fig:NDCG} gives nDCG@k scores for all baseline and proposed models, where $k$ is the required number of recommendations. We use the average number of songs in overall concerts as the upper bound(15 recommendations), while maintaining a simple lower bound(1 recommendation). From these plots, we can evidence that the baseline models give poor performance in giving recommendations for south Indian classical music concerts. Importantly, we can study from Figure~\ref{fig:NDCG} that primarily the proposed ragamAI model give more relevant predictions with even shorter list of recommendations. It is also evident from these plots that structural organization of ragams in Indian classical music concerts play a vital role in building recommendations.

Table~\ref{tab:evolution_statistics} gives the ablation study with multiple combinations in the proposed model. It is evident from these results that each model gives only average performance compared to the proposed model when used alone for the recommendation task. When both features combined as given in the proposed model, we get improved performance.
\begin{table}[!t]
\caption{Ablation study} \centering
\label{tab:evolution_statistics}
\centering

\begin{tabular}{|p{4cm}||p{1.7cm}||p{1.7cm}|}
\hline
\textbf{Model} & \textbf{Precision@15} & \textbf{nDCG@15} \\
\hline
Attention W/O node2vec & 0.1524 & 0.0189 \\
\hline
Attention & 0.2141 & 0.2167 \\
\hline
Concert Embedding & 0.2594 & 0.5398 \\
\hline
\hline
ragamAI & \textbf{0.4875} & \textbf{0.7561} \\
\hline
\end{tabular}
\end{table}
\section{Conclusion and Discussion}
Session-based item recommendation systems have gained many innovations by incorporating deep learning models for variety of applications. Session-based systems can give a precise set of recommendations based on a user's current session activity and their past preferences. In this work we proposed such a model to recommend a set of ragam(music scale in Indian classical music) to perform in a concert, given a set of preferred ragams. Particularly, we proposed a network-based deep learning approach to utilize both hand picked features and structural features of ragam sequences in concerts for the recommendation task. With the experiments, we show that our model can outperform state-of-the-art methods in recommending songs for Indian classical music concerts.

Although this is the first step to merge recommender system with classical music concerts, the model can be extended in several ways in the future: 1. Even though a ragam is an integral part of a song, adding the composer and composition level features to recommend songs benefit and save time for musicians to arrange the concert, 2. Generalize the architecture to support multiple forms of classical music like Western classical music, and 3. Such an extensive model can be adopted in music streaming services like Pandora and Spotify to match the listener preference in the classical music.

\bibliographystyle{IEEEtran} 
{\footnotesize
\bibliography{bibliography}}

\begin{thebibliography}{10}
\providecommand{\url}[1]{#1}
\csname url@samestyle\endcsname
\providecommand{\newblock}{\relax}
\providecommand{\bibinfo}[2]{#2}
\providecommand{\BIBentrySTDinterwordspacing}{\spaceskip=0pt\relax}
\providecommand{\BIBentryALTinterwordstretchfactor}{4}
\providecommand{\BIBentryALTinterwordspacing}{\spaceskip=\fontdimen2\font plus
\BIBentryALTinterwordstretchfactor\fontdimen3\font minus
  \fontdimen4\font\relax}
\providecommand{\BIBforeignlanguage}[2]{{%
\expandafter\ifx\csname l@#1\endcsname\relax
\typeout{** WARNING: IEEEtran.bst: No hyphenation pattern has been}%
\typeout{** loaded for the language `#1'. Using the pattern for}%
\typeout{** the default language instead.}%
\else
\language=\csname l@#1\endcsname
\fi
#2}}
\providecommand{\BIBdecl}{\relax}
\BIBdecl

\bibitem{you2019hierarchical}
J.~You, Y.~Wang, A.~Pal, P.~Eksombatchai, C.~Rosenburg, and J.~Leskovec,
  ``Hierarchical temporal convolutional networks for dynamic recommender
  systems,'' ser. WWW '19, 2019, pp. 2236--2246.

\bibitem{Pichl2017Spotify}
M.~Pichl, E.~Zangerle, and G.~Specht, ``Understanding user-curated playlists on
  spotify: A machine learning approach,'' \emph{Int. J. Multimed. Data Eng.
  Manag.}, vol.~8, no.~4, pp. 44--59, Oct. 2017.

\bibitem{isinkaye2015recommendation}
F.~Isinkaye, Y.~Folajimi, and B.~Ojokoh, ``Recommendation systems: Principles,
  methods and evaluation,'' \emph{Egyptian Informatics Journal}, vol.~16,
  no.~3, pp. 261--273, 2015.

\bibitem{gunawardana2015evaluating}
A.~Gunawardana and G.~Shani, ``Evaluating recommender systems,'' in
  \emph{Recommender systems handbook}.\hskip 1em plus 0.5em minus 0.4em\relax
  Springer, 2015, pp. 265--308.

\bibitem{Bonnin2014automated}
G.~Bonnin and D.~Jannach, ``Automated generation of music playlists: Survey and
  experiments,'' \emph{ACM Comput. Surv.}, vol.~47, no.~2, pp. 26:1--26:35,
  Nov. 2014.

\bibitem{chen2012playlist}
S.~Chen, J.~L. Moore, D.~Turnbull, and T.~Joachims, ``Playlist prediction via
  metric embedding,'' in \emph{ACM SIGKDD}, 2012, pp. 714--722.

\bibitem{Cheng2017exploring}
Z.~Cheng, J.~Shen, L.~Nie, T.-S. Chua, and M.~Kankanhalli, ``Exploring
  user-specific information in music retrieval,'' in \emph{ACM SIGIR}, ser.
  SIGIR '17, 2017, pp. 655--664.

\bibitem{hornung2013evaluating}
T.~Hornung, C.-N. Ziegler, S.~Franz, M.~Przyjaciel-Zablocki, A.~Sch{\"a}tzle,
  and G.~Lausen, ``Evaluating hybrid music recommender systems,'' in
  \emph{Proceedings of the IEEE/WIC/ACM International Joint Conferences on Web
  Intelligence (WI) and Intelligent Agent Technologies (IAT)-Volume 01}, 2013,
  pp. 57--64.

\bibitem{Hidasi2016SessionbasedRW}
B.~Hidasi, A.~Karatzoglou, L.~Baltrunas, and D.~Tikk, ``Session-based
  recommendations with recurrent neural networks,'' \emph{CoRR}, vol.
  abs/1511.06939, 2016.

\bibitem{Jannach2017rnn}
D.~Jannach and M.~Ludewig, ``When recurrent neural networks meet the
  neighborhood for session-based recommendation,'' in \emph{RecSys}, ser.
  RecSys '17, 2017, pp. 306--310.

\bibitem{hidasi2018recurrent}
B.~Hidasi and A.~Karatzoglou, ``Recurrent neural networks with top-k gains for
  session-based recommendations,'' in \emph{Proceedings of the 27th ACM
  International Conference on Information and Knowledge Management}, 2018, pp.
  843--852.

\bibitem{schedl2015tailoring}
M.~Schedl and D.~Hauger, ``Tailoring music recommendations to users by
  considering diversity, mainstreaminess, and novelty,'' in \emph{ACM SIGIR},
  2015, pp. 947--950.

\bibitem{dighe2013swara}
P.~Dighe, H.~Karnick, and B.~Raj, ``Swara histogram based structural analysis
  and identification of indian classical ragas.'' in \emph{ISMIR}, 2013, pp.
  35--40.

\bibitem{kumar2014identifying}
V.~Kumar, H.~Pandya, and C.~Jawahar, ``Identifying ragas in indian music,'' in
  \emph{22nd International Conference on Pattern Recognition}.\hskip 1em plus
  0.5em minus 0.4em\relax IEEE, 2014, pp. 767--772.

\bibitem{gulati2016phrase}
S.~Gulati, J.~Serra, V.~Ishwar, S.~Sent{\"u}rk, and X.~Serra, ``Phrase-based
  r{\=a}ga recognition using vector space modeling,'' in \emph{IEEE
  International Conference on Acoustics, Speech and Signal Processing
  (ICASSP)}, 2016, pp. 66--70.

\bibitem{salamon2012multipitch}
J.~Salamon, S.~Gulati, and X.~Serra, ``A multipitch approach to tonic
  identification in indian classical music,'' in \emph{ISMIR}, 2012.

\bibitem{hu2017diversifying}
L.~Hu, L.~Cao, S.~Wang, G.~Xu, J.~Cao, and Z.~Gu, ``Diversifying personalized
  recommendation with user-session context.'' in \emph{IJCAI}, 2017, pp.
  1858--1864.

\bibitem{wang2018attention}
S.~Wang, L.~Hu, L.~Cao, X.~Huang, D.~Lian, and W.~Liu, ``Attention-based
  transactional context embedding for next-item recommendation,'' in
  \emph{Thirty-Second AAAI Conference on Artificial Intelligence}, 2018.

\bibitem{wu2019session}
S.~Wu, Y.~Tang, Y.~Zhu, L.~Wang, X.~Xie, and T.~Tan, ``Session-based
  recommendation with graph neural networks,'' in \emph{Proceedings of the AAAI
  Conference on Artificial Intelligence}, vol.~33, 2019, pp. 346--353.

\bibitem{aslay2018influence}
C.~Aslay, L.~V. Lakshmanan, W.~Lu, and X.~Xiao, ``Influence maximization in
  online social networks,'' in \emph{ACM WSDM}, 2018, pp. 775--776.

\bibitem{chien2018community}
I.~Chien, C.-Y. Lin, and I.-H. Wang, ``Community detection in hypergraphs:
  Optimal statistical limit and efficient algorithms,'' in \emph{International
  Conference on Artificial Intelligence and Statistics}, 2018, pp. 871--879.

\bibitem{ying2018graph}
R.~Ying, R.~He, K.~Chen, P.~Eksombatchai, W.~L. Hamilton, and J.~Leskovec,
  ``Graph convolutional neural networks for web-scale recommender systems,'' in
  \emph{ACM SIGKDD}, 2018, pp. 974--983.

\bibitem{grover2016node2vec}
A.~Grover and J.~Leskovec, ``node2vec: Scalable feature learning for
  networks,'' in \emph{ACM SIGKDD}, 2016, pp. 855--864.

\bibitem{linden2003amazon}
G.~Linden, B.~Smith, and J.~York, ``Amazon. com recommendations: Item-to-item
  collaborative filtering,'' \emph{IEEE Internet computing}, no.~1, pp. 76--80,
  2003.

\bibitem{rendle2010factorizing}
S.~Rendle, C.~Freudenthaler, and L.~Schmidt-Thieme, ``Factorizing personalized
  markov chains for next-basket recommendation,'' in \emph{ACM WWW}, 2010, pp.
  811--820.

\end{thebibliography}

\end{document}